\documentclass[11pt,a4paper,amssym]{article}

\usepackage{amssymb}
\newcommand{\R}{{\mathbb R}}

\newcommand{\itemi}{\item[{\rm(i)}]}
\newcommand{\itemii}{\item[{\rm(ii)}]}
\newcommand{\itemiii}{\item[{\rm(iii)}]}

\newcommand{\IR}{{\bf I}{\mathbb R}}

\newcommand{\dda}{\mathord{\hbox{\makebox[0pt][l]{\lower .6mm
                           \hbox{$\downarrow$}}$\downarrow$}}}
\newcommand{\dua}{\mathord{\hbox{\makebox[0pt][l]{\raise .6mm
                           \hbox{$\uparrow$}}$\uparrow$}}}

\def\qed{{$\square$}}

\newtheorem{theorem}{Theorem}[section]
\newtheorem{corollary}[theorem]{Corollary}
\newtheorem{lemma}[theorem]{Lemma}
\newtheorem{proposition}[theorem]{Proposition}

\newtheorem{defin}[theorem]{Definition}

\newtheorem{ex}[theorem]{Example}

\newtheorem{prob}{Problem}

\newtheorem{exer}[theorem]{Exercise}

\newenvironment{proof}
  {\begin{trivlist}\item[\hskip\labelsep{\bf Proof}]}{\end{trivlist}}

\title{\bf An Extension of Gleason's Theorem for Quantum Computation}
\author{{\bf Abbas Edalat}\\ Department of Computing, Imperial College London, UK}
\date{}
\begin{document}
\maketitle
\begin{abstract}
  We develop a synthesis of Turing's paradigm of computation and von
  Neumann's quantum logic to serve as a model for quantum computation with recursion,
  such that potentially non-terminating computation can take place, as in a quantum
  Turing machine. This model is based on the extension
  of von Neumann's quantum logic to partial states, defined here as 
  sub-probability measures on the Hilbert space, equipped with the
  natural pointwise partial ordering. The sub-probability
  measures allow a certain probability for the non-termination of the
  computation. We then derive an extension of Gleason's theorem
  and show that, for Hilbert spaces of dimension greater than two, the
  partial order of sub-probability measures is order isomorphic with
  the collection of partial density operators, i.e.\ trace class
  positive operators with trace between zero and one, equipped with the usual partial
  ordering induced from positive operators. We show that the expected value of a bounded observable with respect to a partial state can be defined as a closed bounded interval, which extends the classical definition of expected value.
\end{abstract}

\section{Introduction}
The rigorous mathematical foundation of quantum mechanics is generally
agreed to be based on von Neumann's formulation, which uses the notion
of a state of quantum logic, a probability measure on the collection
of quantum events, i.e., on the closed subspaces of the Hilbert space~\cite{Von55,Var70}.
An observable, discrete or continuous, is defined as a mapping from the
Borel subsets of the real line to the collection of states of the
quantum logic. These two notions are then used to derive the expected
value of an observable, which provides a consistent quantum theory
treating discrete and continuous observables in a uniform manner.
Gleason's fundamental theorem shows that, for Hilbert spaces of dimension greater
than two, the states of logic are in one to one correspondence with
density operators, i.e. trace class operators on the Hilbert space
with trace one, and thus allows us to work with the more convenient
density operators instead of the probability measures on the collection
of closed subspaces of the Hilbert space.

In this paper, we aim to synthesize von
Neumann's formulation of quantum mechanics with Turing's paradigm of
computation, in which classical iteration or recursion plays a foundational role.

All standard programming languages allow for non-trivial recursion, a
fundamental feature of computation which can potentially result in the
non-termination of programs. In a pioneering paper in 1936, Alan
Turing proved that it is not decidable in general if a program
terminates, i.e.\ one cannot in general determine, in
a finite amount of time, if a given program terminates on a given
input~\cite{Tur36,Tur37}. This extremely important result in computer
science has since been called ``the halting problem''. It is a basic
consequence of the halting problem that in probabilistic programs,
when one computes the probability of the outcome of an event as in
quantum computation, sub-probability measures, rather than probability measures, are used to model the probabilistic results: the total probabilities of the definite outcome add up to a number less than one, allowing a certain probability for the non-termination of the program; see for example~\cite{Koz81}. 

In order to reconcile von Neumann's quantum logic with the halting
problem, we therefore consider sub-probability measures on quantum
events, i.e.\ measures $\mu$ on the closed subspaces of the Hilbert
space ${\cal H}$ such that $0\leq \mu({\cal H})\leq 1$, which we call {\em
  partial states of the quantum logic}. There is a natural partial
order on partial states: $\mu_1\leq \mu_2$ if $\mu_1(A)\leq \mu_2(A)$
for all closed subspace $A$ of ${\cal H}$, i.e., for any quantum event
the probability of outcome with respect to $\mu_2$ is at least as much
as that of $\mu_1$, which implies that the probability of
non-termination for $\mu_2$ is at most that of $\mu_1$. This partial
order is complete in that any increasing chain has a supremum, more
generally it is directed complete in the sense that any directed set
has a supremum. Recall that a non-empty subset of a partially ordered
set is {\em directed} if for any two elements in the subset there is
an element of the subset above both; thus a directed subset is a
generalization of an increasing chain in a partial order. A directed
set of partial states represents a consistent set of computations: for
any pair of partial states in the set, there exists a partial state in
the set that assigns to each quantum event a probability at least as
great as either of those assigned to the quantum event by the two partial
states.

We then consider trace class positive operators with trace
between zero and one, which we call {\em partial density operators},
equipped with the partial ordering given by $B\leq A$ if $A-B$ is a
positive operator. This partial order is also directed complete, i.e., any directed subset has a supremum. 
We show that, for Hilbert
spaces of dimension greater than two, Gleason's onto map from the
collection of density operators to that of states of quantum logic
extends to an order isomorphism between the directed complete partial
order of partial density operators and that of the partial states of
the logic, preserving the supremum of directed subsets. Any directed
complete partial order is equipped with a natural $T_0$ topology, the
so-called {\em Scott topology}, which corresponds to the convergence
of a directed set to its supremum. The extended Gleason map is then
continuous with respect to the Scott topologies of the directed
complete partial orders of the partial density operators and the
partial states of the quantum logic. All in all, this shows that
partial density operators provide a convenient model for quantum
computation with recursion.  The partial order of partial density operators
has already been used in~\cite{Sel03} to develop a functional
programming language for quantum computation. 
In a typical recursive computation, the
partial density operator in each loop of iteration increases in the
partial order and in the limit one obtains the supremum of this
increasing chain of partial density operators, which can indeed be a
usual density operator; see~\cite[Section 5.5]{Sel03} for an example.

Finally, we consider any bounded observable, with a discrete or continuous spectrum, given as a map from the Borel subsets of the real line to the lattice of closed subsets of the Hilbert space. We define the expected value of the observable with respect to a partial state as a compact real interval and show that, for a fixed bounded observable, the expected value as a function from the directed complete partial order of partial states to the directed complete partial order of the collection of non-empty compact intervals ordered by reverse inclusion is Scott continuous.  By the spectral theory of operators, the bounded observables are in one-to-one correspondence with bounded self-adjoint operators on the Hilbert space. We thus obtain the expected value of a bounded self adjoint operator with respect to a partial density matrix and show that for a fixed partial density operator, the expected value, as an interval-valued map, depends linearly on commuting bounded self adjoint operators.
\section{Partial states of quantum logic}
To fix our notations, we briefly review von Neumann's formulation of the states of quantum logic. Let ${\cal H}$ be a separable (finite or infinite dimensional) Hilbert space over the complex numbers. We denote the set of closed subspaces of ${\cal H}$ partially ordered by inclusion by $S({\cal H})$. A closed subspace is also referred to as a {\em quantum event}. Then $S({\cal H})$ is a non-distributive complete lattice with $A\wedge B=A\cap B$ and $A\vee B= \mbox{span}\{A,B\}$, the subspace generated by $A$ and $B$. 

A {\em probability measure} on $S({\cal H})$ is a mapping $p:S({\cal H})\to [0,1]$ such that
\begin{itemize}
\itemi $p(\{0\})=0$,
\itemii $p(H)=1$,
\itemiii $p(\bigvee_{n\geq 0}A_n)=\sum_{n\geq 0}p(A_n)$, for any sequence $A_n$ of mutually orthogonal subspaces $A_n$.
\end{itemize}

A {\em state} of the quantum logic or of quantum events is a probability measure on $S({\cal H})$; it gives the probability of each event $A\in S({\cal H})$ and thus all the information required to compute any physical property of the system. We now consider any quantum programming language, which computes probabilities of quantum events and is allowed to use recursion. A {\em partial state} of the quantum logic is defined as a {\em subprobability measure} of $S({\cal H})$, i.e., as a mapping $p:S({\cal H})\to [0,1]$ such that
\begin{itemize}
\itemi $p(\{0\})=0$,
\itemii $p(H)\leq 1$,
\itemiii $p(\bigvee_{n\geq 0}A_n)=\sum_{n\geq 0}p(A_n)$, for any sequence $A_n$ of mutually orthogonal subspaces.
\end{itemize}
The number $1-p(H)\geq 0$ is the probability of {\em non-termination}
of the program or the computation process. We note that this number,
and hence the probability of termination, can be non-computable when
the Hilbert space ${\cal H}$ is infinite dimensional.

In fact, for an orthonormal basis $(e_i)_{i\geq 0}$ of ${\cal H}$ and any non-computable number $a$ between $0$ and $1$, there is a subprobability measure $p$ on ${\cal H}$ with $p({\cal H})=a$, such that $p$ assigns a rational number to any finite subspace generated by $(e_i)_{i\geq 0}$. Such $p$ can be defined as follows. We consider the binary expansion of $a$:
\[a=\sum_{i\geq 0} \frac{a_i}{2^{i+1}},\]
where $a_i\in\{0,1\}$ for $i\geq 0$.and define the partial state $p$ by its values on the basis vectors as $p(\mbox{span}\{e_i\})= \frac{a_i}{2^{i+1}}$. Then $p$ assigns a rational number to any finite dimensional subspace generated by the orthonormal basis $(e_i)_{i\geq 0}$ and we have $p({\cal H})=a$.

There is a natural notion
of partial order on partial states. For two partial states, we define
$p_1\leq p_2$, if for all quantum events $A\in S({\cal H})$, we have
$p_1(A)\leq p_2(A)$, i.e., $p_2$ gives more probability to any quantum
event and is more likely to terminate compared to $p_1$. We need the following lemma, which is a generalization of a corresponding well-known result for double sequences of real numbers.
\begin{lemma}
If $(a_{ij})_{i\in I,j\in J}$ is a double indexed bounded subset of real numbers, which is directed in each index, then 
\[\sup_{i\in I}\sup_{j\in J}a_{ij}=\sup_{j\in J}\sup_{i\in I}a_{ij}.\]
\end{lemma}

\begin{proposition}
The partial order on partial states is directed complete.
\end{proposition}
\begin{proof}
Given a directed subset $(p_i)_{i\in I}$ of partial states, consider the map $p:S({\cal H})\to [0,1]$ given by $p=\sup_{i\in I}p_i$, i.e., $p(A)=\sup_{i\in I} p_i(A)$ for any Borel subset $A\subseteq \R$. Clearly $p(\{0\})=0$ and $p(H)\leq 1$. Suppose $(A_j)_{j\geq 0}$ is a sequence of mutually orthogonal subspaces. For each $i\in I$ we have by assumption: $p_i(\bigvee_{n\geq 0}A_n)=\sum_{n\geq 0}p_i(A_n)$. Let $B_{ij}=\sum_{n=0}^jp_i(A_n)$ be the partial sum of the infinite series of nonnegative terms $p_i(\bigvee_{n\geq 0}A_n)$. Then the bounded double indexed subset $(B_{ij})_{i\in I,j\geq 0}$ of non-negative numbers is directed in each index. It follows from the lemma that we can change the order of taking suprema in $B_{ij}$ and thus:

\[\begin{array}{llll}  p(\bigvee_{n\geq 0}A_n)&=&\sup_{i\in I}p_i(\bigvee_{n\geq 0}A_n)\\
                                              &=&\sup_{i\in I}\sup_{j\geq 0} B_{ij}\\
                                              &=&\sup_{j\geq 0}\sup_{i\in I} B_{ij}\\
                                             &=&\sup_{j\geq 0}\sup_{i\in I}\sum_{n=0}^jp_i(A_n)\\
                                            &=&\sup_{j\geq 0}\sum_{n=0}^jp(A_n)\\
                                              &=& \sum_{n\geq 0}p(A_n)\\
\end{array}\]
It follows that $p$ is a partial state. It is easy to check that $p$ is indeed the supremum of the subset $(p_i)_{i\in I}$.~\qed
\end{proof}
We denote the complete partial order of partial states (subprobability measures) by $M(S({\cal H}))$ and the set of states (probability measures) by $M^1(S({\cal H}))$.
\section{Extension of Gleason's theorem}
Consider the set $D^1$ of density operators, in other words, positive linear operators $f:H\to H$ of trace class with trace equal to one, i.e., $\langle x|fx\rangle\geq 0$ for all $x\in H$, which we write as $f\geq 0$, and $\mbox{tr}(f)=\sum_{i\geq 0}\langle e_i|fe_i\rangle=1$ for an orthonormal basis $(e_i)_{i\geq 0}$ of ${\cal H}$. In fact, the trace of a positive operator is independent of the orthonormal basis. Consider the linear map $G:D^1\to M^1(S({\cal H}))$ given by $G(f):S({\cal H})\to [0,1]$ with $G(f)(K)=\mbox{tr}(P^Kf)$, where $P^K$ is the orthogonal projection onto the closed subspace $K\in S({\cal H})$. The following celebrated theorem was assumed by von Neumann, then conjectured by Mackey and finally proved by Gleason. A more elementary proof was later provided in~\cite{CKM85}.
\begin{theorem}\cite{Gle57}
The map $G$ is onto if the dimension of ${\cal H}$ is greater than two.~\qed
\end{theorem}
In order to obtain an extension of Gleason's theorem for partial states, we consider the notion of {partial density operators}. A positive linear operator $f:H\to H$ of trace class is a {\em partial density operator} if $\mbox{tr}(f)\leq 1$. Partial density operators are equipped with the usual partial ordering of operators, namely $g\leq f$ if $f-g$ is a positive operator, written as: $f-g\geq 0$. Let $D$ be the partial order of partial density operators. We use the following result of Vigier.
\begin{theorem}~\cite[page 51]{Thi79}
Every norm bounded increasing filter of operators has a supremum.
\end{theorem}
\begin{corollary}
The partially ordered set of partial density operators is directed complete.
\end{corollary}
\begin{proof}
A directed set in $D$ is an increasing filter with respect to the partial order. Since for any operator $f\in D$, we have $\|f\|\leq \mbox{tr}(f)\leq 1$, it follows that any directed set in $D$ is bounded with respect to the operator norm and the result follows from the above theorem.~\qed
\end{proof}

Recall that for directed complete partially ordered sets $R$ and $S$, a mapping $h:R\to S$ is an {\em order isomorphism} if $h$ is one-to-one, onto, monotone ($h(x)\leq h(y)$ if $x\leq y$) and reflects the partial order ($x\leq y$ if $h(x)\leq h(y)$). We note that any order reflecting map is necessarily one-to-one: $h(a)=h(b)$ implies $h(a)\leq h(b)$ and $h(b)\leq h(a)$, from which it follows that $a\leq b$ and $b\leq a$, i.e., $a=b$.

\begin{proposition}If $h:R\to S$ is an order isomorphism, then it preserves the supremum of directed subsets, i.e., $\sup_{i\in I}h(a_i)=h(\sup_{i\in I}a_i)$ for any directed subset $(a_i)_{i\in I}$. 
\end{proposition}
\begin{proof}
By monotonicity of $h$ we have: $\sup_{i\in I}h(a_i)\leq h(\sup_{i\in I}a_i)$. Since $h$ is onto, there exists $a\in R$ such that $h(a)=\sup_{i\in I}h(a_i)$. Thus, for each $i\in I$, we have: $h(a_i)\leq h(a))\leq h(\sup_{i\in I}a_i)$.  By order reflection, it follows that for $i\in I$, we have: $a_i\leq a\leq \sup_{i\in I}a_i$. Hence, $a$ is an upper bound of the directed subset $(a_i)_{i\in I}$ and therefore $\sup_{i\in I}a_i\leq a$, which implies $a=\sup_{i\in I}a_i$.~\qed  \end{proof}

We will now state and prove the extension of Gleason's map to partial density operators and partial states. 
\begin{theorem}
For Hilbert spaces of dimension greater than two, the directed complete partially ordered sets of the partial density operators and the partial states are order isomorphic.
\end{theorem}
\begin{proof}
Consider the linear extension $G:D\to M(S({\cal H}))$ given by $G(f):S({\cal H})\to [0,1]$ with $G(f)(K)=\mbox{tr}(P^Kf)$, where $P^K$ is as before the orthogonal projection onto the closed subspace $K\in S({\cal H})$. To show that $G$ is monotone, suppose $g\leq f$ and assume $K\in S({\cal H})$. Let $(e_i)_{i\geq 0}$ be a complete orthonormal set of eigenvectors of $P^K$. Then,
\[\begin{array}{lll}
\mbox{tr}P^K(f-g)&=&\sum_{i\geq 0}\langle e_i\mid P^K(f-g)e_i\rangle\\
&=&\sum_{i\geq 0}\langle P^Ke_i\mid (f-g)e_i\rangle\\
&=&\sum_{P^Ke_i=e_i}\langle e_i\mid (f-g)e_i\rangle\geq 0.\\
\end{array}\]
To show that $G$ is order reflecting, assume $G(g)\leq G(f)$. If, for some $x\in H$ with $\|x\|=1$ we have $\langle x\mid (f-g)x\rangle<0$, then by taking $K$ to be the subspace generated by $x$ and using an orthonormal basis $(e_i)_{i\geq 0}$ with $e_0=x$, we obtain:
\[\begin{array}{lll}
G(f)-G(g)&=&\mbox{tr} P^K(f-g)\\
&=&\sum_{i\geq 0}\langle e_i\mid P^K(f-g)e_i\rangle\\
&=&\sum_{i\geq 0}\langle P^Ke_i\mid (f-g)e_i\rangle\\
&=&\langle x\mid (f-g)x\rangle<0,\\
\end{array}\]
which gives a contradiction. This proves order reflectivity, from which it follows that $G$ is one-to-one. The surjectivity of $G$ from Gleason's theorem by noting that for $0\leq r\leq 1$ and all $f\in D$, we have: $G(rf)=rG(f)$.~\qed
\end{proof}
We note that the {\em Scott topology} of a directed complete partial order $A$ is given as follows. The open subsets of the Scott topology are those subsets $O\subseteq A$ such that (i) $O$ is upward closed, i.e., $x\in O$ and $x\leq y$ implies $y\in O$, and (ii) $O$ is inaccessible by directed subsets, i.e., whenever $\sup_{i\in I}x_i\in O$ for a directed subset $(x_i)_{i\in I}$, then there exists $i\in I$ with $x_i\in O$. The Scott topology on a directed complete partial order is in general $T_0$ and is the canonical topology with respect to which a directed set (net) converges to its supremum. In fact, in analogy with continuous maps between metric spaces, we have the following property: A mapping $h:A\to B$ between directed complete partial orders is continuous with respect to the Scott topologies of $A$ and $B$ if and only if $f$ is monotone and preserves the supremum of directed subsets~\cite{AJ94}. We have thus obtained canonical topologies on $D$ and $M(S({\cal H}))$ with respect to which the one-to-one correspondence between the partial density operators and the partial states is in fact a homeomorphism:
\begin{corollary}
The map $G:D\to M(S({\cal H})$ is continuous with respect to the Scott topology.~\qed
\end{corollary}
\section{Expected values of observables}
Recall~\cite{Var70} that an observable is a map $r:{\cal B}(\R)\to S({\cal H}))$, where ${\cal B}(\R)$ is the collection of Borel subsets of the real line, such that:
\begin{itemize}
\itemi $r(\emptyset)=0$ and $r(\R)={\cal H}$.
\itemii $r(U)$ and $r(V)$ are orthogonal subspaces if $U\cap V=\emptyset$.
\itemiii $r(\bigcup_{n\geq 0}U_n)=\bigvee_{n\geq 0}r(U_n)$ for any sequence $(U_n)_{n\geq 0}$ of Borel subsets.
\end{itemize}
The interpretation of the map $r$ as an observable is that for the quantum event $r(U)\in S({\cal H})$, with $U\subseteq \R$ a Borel subset, the observable takes its values in $U$. The observable $r$ is {\em bounded} if there exists $a>0$ such that $r([-a,a])={\cal H}$. The spectrum $\mbox{Spec}(r)$ of $r$ is the intersection of all closed subsets $C\subseteq \R$ such that $r(C)={\cal H}$. By the spectral theory of operators, there is a one-to-one correspondence between (bounded) observables and (bounded) self-adjoint operators on ${\cal H}$~\cite[pages 235, 263]{RS80}, which are the standard representation of observables in quantum mechanics.

If $p:S({\cal H})\to [0,1]$ is a state, then the composition $Q^r_p=p\circ r:{\cal B}(\R)\to [0,1]$ with $Q^r_p(U)=p(r(U))$ is a probability distribution on $\R$ and the {\em expected value} of $r$ in the state $p$ is defined by 
\[E(r\mid p)=\int^\infty_{-\infty}t\,dQ^r_p(t),\]
when the integral exists. If $r$ is a bounded observable and $p=G(f)$ for some $f\in D^1$, then the expected value of $r$ with respect to $p$ exists and is given by
\[E(r\mid p)=\mbox{tr}(A_rf),\]
where $A_r$ is the unique bounded self adjoint operator corresponding to $r$~\cite[page 61]{Var70}.

The following question now naturally arises. Given an observable, can
we define the expected value of a partial state, extending the classical definition of the expected value with respect to a state? We will show here that we can do this in a satisfactory way for the case of bounded observables by defining the expected value with respect to a partial state as an interval rather than a real number.

Consider the observable $r$ with respect to a partial state $p:S({\cal H})\to [0,1]$. Let $m=\inf\mbox{Spec}(r)$ and $M=\sup\mbox{Spec}(r)$. The composition $Q^r_p=p\circ r:{\cal B}(\R)\to [0,1]$ with $Q^r_p(U)=p(r(U))$ is a subprobability distribution on $\R$. We put
\[E_0(r\mid p)=\int^M_{m}t\,dQ^r_p(t),\]
when the integral exists. When $r$ is bounded, we define the expected value of $r$ with respect to the partial state $p$ as the closed bounded interval: 
\begin{equation}\label{ex}E(r|p)=E_0(r\mid p)+(1-Q^r_p(\R))[m,M],\end{equation}
where we have used the standard notation in interval arithmetic: $k+[a,b]=[k+a,k+b]$ for a real number $k$ and a real interval $[a,b]$. 
The second term in Equation~\ref{ex} says that the missing probability, i.e.\ $(1-Q^r_p(\R))$, is known to be distributed on $[m,M]$ but, at this stage of computation, its precise distribution on this interval remains indeterminate. This uncertainty explains why the expected value with respect to a partial state is an interval rather than a real number. For a non-empty compact interval $a$, we use the notation $a=[\underline{a},\overline{a}]$ for the left and right end points of the interval. 
\begin{proposition}\label{expect-monotone}
Suppose $r$ is a bounded observable. If $p\leq q$, then $E(r|p)\supseteq E(r|q)$.
\end{proposition}
\begin{proof}
Let $\mu=Q^r_q-Q^r_p$. Then by assumption $\mu$ is a nonnegative measure with support in $[m,M]$. Thus:
\[\intop^M_mt\,d\mu(t)\geq \mu([m,M])m,\]
and, hence, 
\[\int^M_mt\,dQ^r_q(t)-\int^M_mt\,dQ^r_p(t)\geq (Q^r_q(\R)-Q^r_q(\R))m.\]
It follows that $\underline{E(r|p)}\leq\underline{E(r|q)}$. Similarly, $\overline{E(r|q)}\leq\overline{E(r|p)}$.~\qed
\end{proof}
We now aim to show that our notion of expected value has the required limiting properties.
\begin{lemma}
Suppose $r$ is any observable. If $(p_{i})_{i\in I}$ is a directed set of partial states with $p=\sup_{i\in I}p_i$ and $k:\R\to \R$ is any measurable function such that $\int_{-\infty}^\infty k\,dQ^r_p$ exists, then 
\[\int_{-\infty}^\infty k\,dQ^r_p=\lim_{i\in I}\int_{-\infty}^\infty k\,dQ^r_{p_i},\]
where the limit is understood as the limit of the net $(\int_{-\infty}^\infty k\,dQ^r_{p_i})_{i\in I}$.

\end{lemma} 
\begin{proof}
Let $\mu_i=Q^r_{p_i}$ and $\mu=Q^r_{p}$.  Let $h:\R\to \R$ be any simple function, i.e., $h=\sum_{j=1}^n\chi_{B_j}$ for some measurable subsets $B_i\subseteq \R$ ($1\leq j\leq n$), where $\chi_B$ is the characteristic function of $B$. Then, 
\[\begin{array}{lll}
\sup_{i\in I}\int_{-\infty}^{\infty}h\,d\mu_i&=&\sup_{i\in I}\int_{-\infty}^{\infty}\sum_{j=1}^n\chi_{B_j}\,d\mu_i\\[1ex]
&=&\sup_{i\in I}\sum_{j=1}^n\mu_i({B_j})\\[1ex]
&=&\sum_{j=1}^n\mu({B_j})\\[1ex] 
 &=&\int_{-\infty}^{\infty}\sum_{j=1}^n\chi_{B_j}\,d\mu\\[1ex]
 &=&\int_{-\infty}^{\infty}h\,d\mu.
\end{array}\]
Assume first that $k$ is nonnegative. Since the set of simple functions $h$ with $0\leq h\leq k$ is directed, from the above we obtain:
\[\begin{array}{lll}\intop_{-\infty}^\infty k\,d\mu&=&\sup\{\intop_{-\infty}^\infty h\,d\mu\mid \mbox{nonnegative simple }h\leq k\}\\
&=&\sup\{\sup_{i\in I}\intop_{-\infty}^\infty h\,d\mu_i\mid \mbox{nonnegative simple }h\leq k\}\\
&=&\sup_{i\in I}\sup\{\intop_{-\infty}^\infty h\,d\mu_i\mid \mbox{nonnegative simple }h\leq k\}\\
&=&\sup_{i\in I}\intop_{-\infty}^\infty k\,d\mu_i\\
&=&\lim_{i\in I}\intop_{-\infty}^\infty k\,d\mu_i.\\
\end{array}\]
This establishes the result for a positive function $k$.  More
generally, we split $k$ into its positive and negative parts and put
$k^+=\max(k,0)$ and $k^-=\max(-k,0)$, where $0$ denotes the constant
function with value zero. Since $\int_{-\infty}^\infty k\,dQ^r_p$
exists, it follows that both $\int_{-\infty}^\infty k^+\,dQ^r_p$
and$\int_{-\infty}^\infty k^-\,dQ^r_p$ exist as well. From
$k=k^+-k^-$, we obtain, using the result for a positive function:
\[\begin{array}{lll}
\intop_{-\infty}^\infty k\,d\mu&=&\intop_{-\infty}^\infty k^+\,d\mu-\intop_{-\infty}^\infty k^-\,d\mu\\
 &=&\lim_{i\in I}\intop_{-\infty}^\infty k^+\,d\mu_i-\lim_{i\in I}\intop_{-\infty}^\infty k^-\,d\mu_i\\
&=&\lim_{i\in I}\intop_{-\infty}^\infty k^+-k^-\,d\mu_i\\
&=&\lim_{i\in I}\intop_{-\infty}^\infty k\,d\mu_i.\qquad\mbox{\qed}\\
\end{array}\]
\end{proof}
\begin{corollary}\label{converge-expect0}
If $(p_{i})_{i\in I}$ is a directed set of partial states, then for any observable $r$ such that $E_0(r\mid \sup_{i\in I} p_i)$ exists, we have:
\[\sup_{i\in I} E_0(r\mid p_i)=E_0(r\mid \sup_{i\in I} p_i).\]
\end{corollary}

We can now show that the expected value of an observable with respect to a partial state has the desired limiting properties. Note that the collection of non-empty compact intervals of the real line partially ordered by reverse inclusion, denoted by $\IR$, is a directed complete partial order, in which the supremum of a directed set is the intersection of the compact intervals represented by the directed set.
\begin{theorem}
For a given bounded observable $r$, the expected value map $E(r|.):M(S({\cal H}))\to \IR$ is Scott continuous.
\end{theorem}
\begin{proof}We have already checked in Proposition~\ref{expect-monotone} that the map $E(r|.)$ is monotone. Let $(p_{i})_{i\in I}$ be a directed set of partial states with $p=\sup_{i\in I}p_i$. Then, by Corollary~\ref{converge-expect0}, we have:
\[\begin{array}{lll}\lim_{i\in I}\underline{E(r|p_i)}&=&\lim_{i\in I} (E_0(r|p_i)+m(1-Q^r_{p_i}(\R)))\\
&=&\lim_{i\in I} E_0(r|p_i)+m(1-\lim_{i\in I} Q^r_{p_i}(\R)))\\
&=&E_0(r|p)+m(1-Q^r_{p}(\R)))=\underline{E(r|p)}.\\
\end{array}\]
Similarly, $\lim_{i\in I}\overline{E(r|p_i)}=\overline{E(r|p)}$ and thus: $\bigcap_{i\in I}E(r|p_i)=E(r|p)$.~\qed
\end{proof}
The representation of observables as functions of type ${\cal B}(\R)\to S({\cal H})$ has serious drawbacks when studying the calculus of observables. For example, representing a linear combination of commuting observables and computing its expected value in this setting are far from straightforward; see~\cite[pages 125, 163]{Var70}. In fact, for this purpose, it is far simpler to represent observables as self adjoint operators on the Hilbert space. 
\begin{proposition}
If $A_r$ is the self adjoint operator corresponding to the bounded observable $r$ and if $p=G(f)$ is a partial state with $f$ a partial density operator, then we have:
\[E_0(r|p)=\mbox{tr}(A_rf).\]
\end{proposition}
\begin{proof}
In case, $p$ is a state, and hence $f$ a density operator, the result follows from~\cite[theorem 7.24]{Var70}. If $p=0$ is the trivial partial state, then $f$ is the zero operator and we have $E_0(r|p)=\mbox{tr}(A_rf)=0$. Otherwise k=$p({\cal H})>0$ and $p/k$ is a state. By linearity of $G$, we have: $G(f/k)=p/k$. Now the linearity of integration and the trace function implies:
\[E_0(r|p)=kE_0(r|p/k)=k\mbox{tr}(A_rf/k)=\mbox{tr}(A_rf).\qquad\mbox{\qed}\]
\end{proof}
\begin{corollary}With the above notations:
\[E(r|p)=\mbox{tr}(A_rf)+(1-\mbox{tr}(f))[m,M].\qquad\mbox{\qed}\]
\end{corollary}
Since every partial density operator $f$ corresponds to a unique partial state $G(f)$, we can now write the expected value of a bounded self adjoint operator $A$ with respect to a partial density operator $f$ as
\[{\cal E}(A|f)=\mbox{tr}(Af)+(1-\mbox{tr}(f))[m,M],\]
where $m=\inf\mbox{Spec}(A)$ and $M=\sup\mbox{Spec}(A)$. 
Finally, we deduce that, as an interval valued map, the expected value with respect to a partial state depends linearly on commuting observables. For a bounded self adjoint operator $C$, we write: $m_C=\inf \mbox{Spec}(C)$ and $M_C=\sup \mbox{Spec}(C)$. Then $[m_{kC},M_{kC}]=k[m_C,M_C]$, for any real number $k$, and, for commuting bounded observables $A$ and $B$, we have: $m_{A+B}=m_A+m_B$ and $M_{A+B}=M_A+M_B$, i.e.,\[[m_{A+B},M_{A+B}]=[m_A,M_A]+[m_B,M_B].\] 
\begin{proposition}
If $A$ and $B$ are commuting bounded self adjoint operators, $k$ and $\ell$ real numbers, and $f$ a partial density operator, then 
\[{\cal E}(kA+\ell B|f)=k{\cal E}(A|f)+\ell{\cal E}(B|f).\]
\end{proposition}
\begin{proof}
We have: $[m_{kA+\ell B},M_{kA+\ell B}]=k[m_A,M_A]+\ell [m_B,M_B]$.
Hence,
\[\begin{array}{lll}{\cal E}(kA+\ell B|f)&=&\mbox{tr}((kA+\ell B)f)+(1-\mbox{tr}(f))[m_{kA+\ell B},M_{kA+\ell B}]\\
&=&k\mbox{tr}(Af)+\ell\mbox{tr}(Bf)+k(1-\mbox{tr}(f))[m_A,M_A]+\ell (1-\mbox{tr}(f))[m_B,M_B]\\
&=&k{\cal E}(A|f)+\ell{\cal E}(B|f).\qquad \mbox{\qed}\\\end{array}\]
\end{proof}
One can also compute the expected value of various functions of an observable. For example, for a bounded self-adjoint operator $A$ and a partial density operator $f$ we have:
\[{\cal E}(A^2|f)=\mbox{tr}(A^2f)+(1-\mbox{tr}(f))[m_{A^2},M_{A^2}],\]
where $m_{A^2}=k^2$ and  $M_{A^2}=K^2$, with $k=\inf\{|a| : a\in \mbox{Spec}(A)\}$ and  $K=\sup\{|a| : a\in \mbox{Spec}(A)\}$.
\section*{Acknowledgements}I would like to thank Peter Selinger and Iain Stewart for valuable feedback on the first draft of this paper.

\end{document}